\documentclass[12pt,pdflatex]{iopart}

\usepackage{iopams}
\usepackage{graphicx}
\usepackage{appendix}
\usepackage{wrapfig}
\usepackage{color}
\usepackage{soul}
\newcommand{\bra}[1]{ \langle #1 |}
\newcommand{\ket}[1]{| #1 \rangle}

\usepackage{amssymb}
\usepackage{upgreek}

\usepackage{array}

\usepackage{multirow}

\begin{document}

\title[]{Magneto-quantum-nanomechanics: ultra-high Q levitated mechanical oscillators}

\author{M. Cirio$^{1,\dag}$\footnote[0]{$^{\dag}$ Author to whom any
correspondence should be addressed.}, J. Twamley$^{1}$, G.K. Brennen$^{1}$.}

\address{$^1$Centre for Engineered Quantum Systems, Department of Physics and Astronomy, Macquarie University, North Ryde, NSW 2109, Australia}

\ead{mauro.cirio@students.mq.edu.au}


\begin{abstract}
{\bf Engineering nano-mechanical quantum systems possessing ultra-long motional coherence times allow for applications in ultra-sensitive quantum sensing, motional quantum memories and motional interfaces between other carriers of quantum information such as photons, quantum dots and superconducting systems. To achieve ultra-high motional $Q$ one must work hard to remove all forms of motional noise and heating. We examine a magneto-nanomechanical quantum system that consists of a 3D arrangement of miniature superconducting loops which is stably levitated in a static inhomogenous magnetic field. The resulting motional $Q$ is limited by the tiny decay of the supercurrent in the loops and may reach up to $Q\sim 10^{10}$. We examine the classical and quantum dynamics of the levitating superconducting system and prove that it is stably trapped and can achieve motional oscillation frequencies of several tens of MHz. By inductively coupling this levitating object to a nearby flux qubit we further show that by driving the qubit one can cool the motion of the levitated object and in the case of resonance, this can cool the vertical motion of the object close to itÕs ground state.}

\end{abstract}

\maketitle

{\pagestyle{plain} \tableofcontents }

\section{Introduction}



Recently there has been considerable effort towards mapping the boundary between the classical and the quantum world by exploring the physics of mesoscopic and macroscopic mechanical systems. From an applications point of view, as precision measurement of position and acceleration generally  involve some kind of motion, the necessity of building smaller and more sensitive devices has required a more careful exploration of the classical-quantum limit.

The possibility to couple, control and measure micro-mechanical motion in a wide range of different physical systems leads to new experimental applications in different fields such as measuring forces between individual biomolecules ~\cite{Bustamante,Friedsam,Benoit}, magnetic forces from single spins ~\cite{Rugar},  perturbations due to the mass fluctuations involving single atoms and molecules ~\cite{Ekinci}, pressure ~\cite{Burns1} and acceleration ~\cite{Burns1},  fundamental constants ~\cite{Schilling}, small changes in electrical charge ~\cite{Cleland}, gravitational wave detection ~\cite{Bocko}, as well as applications in quantum computation ~\cite{Nakamura}, quantum optics ~\cite{Brune} and condensed matter physics ~\cite{Clarke,Silvestrini}. 

Observing any quantum properties of a mechanical system is a challenge.  Under typical conditions, energy losses, thermal noise and decoherence processes make it impossible to observe any motional quantum effects. To observe quantum mechanical motional effects the system has to be close enough to its ground state and it has to preserve this quantum coherence  for a reasonable amount of time. This leads to the necessity of engineering  ultra-low dissipative systems  (which, in oscillating systems, is measured by the quality factor $Q$ representing the energy lost per cycle). To achieve this one must engineer a system which is mechanically isolated from it's surroundings to an extreme level. On the other hand one must  also find a way to cool down the motion close to its motional ground state which necessities coupling that system to another in order to dump entropy. Numerous nano-mechanical oscillating systems have been recently studied such as  cavity optomechanical experiments employing cantilevers \cite{Kleckner},  micro-mirrors \cite{Arcizet,Gigan}, micro-cavities \cite{Kippenberg2, Schliesser}, nano-membranes \cite{Schliesser2},  macroscopic mirror modes \cite{Corbitt} and optically levitated nanospheres \cite{Chang} (see ~\cite{Kippenberg}). As shown in Ref. \cite{OConnell}, it has been possible to create and control quantum states but, except in a few cases, reaching large $Q$ for nano to microscopic sized motional devices is still an open problem. In fact, a mechanical oscillator usually involves many degrees of freedom coupled together while we are interested in the quantum behaviour of one of them: the centre of mass motion.

In our work, we present a theoretical model for a mesoscopic mechanical oscillator. It is  inherently non-dissipative (with a estimated $Q\sim 10^{10}$). It consists of a cluster of superconducting loops levitating in vacuum and its motion can be completely characterized by the six degrees of freedom of a rigid body. Controlling the system is possible through an interaction with a nearby flux qubit which is tuned  and driven to allow the levitated structure to be cooled close to it's motional ground state.\\
We will show how to cool the center of mass motion along one degree of freedom to the ground state in an efficient way by coupling the loop inductively to a flux qubit.


\section{Toy Model}
In order to achieve a qualitative understanding of the dynamics of the levitated system presented here we will start by studying a simple 2-dimensional toy model \cite{Romer} which describes  how it is possible to take advantage of the Meissner effect to obtain a stable magnetically levitated mechanical system.
\begin{figure}[h!]
    \begin{center}
\includegraphics[scale=0.45]{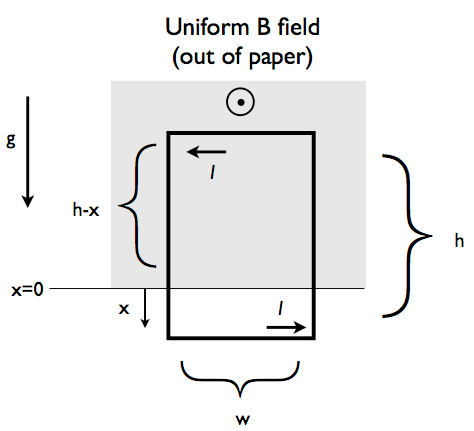}
\end{center}
\caption{A metallic wire loop in a discontinuous magnetic field. When released to fall under gravity the change in flux caused by the loops' motion  will induce current flow in the wires which will oppose the downwards motion and the loop's height will undergo harmonic oscillation. For wires made from normal metals the oscillations are damped  due to resistance within the wires and the loop will eventually fall but for superconducting wires the oscillations will continue for many periods yielding a motional oscillator with extremely high motional $Q$ factor.}
\label{Romer}
\end{figure}
For this toy model, we consider a wire loop of superconducting material oriented in a vertical plane (Fig. ~\ref{Romer}) and threaded by a discontinuous horizontal magnetic field which is uniform and non-zero above a certain height and which vanishes below. Initially the loop is at rest at a position where it crosses the magnetic field discontinuity line.

When the loop falls under the force of gravity, the changing magnetic flux within the loop will induce circulating currents which generate Lorentz forces that counteract the downwards motion of the loop. If the loop is made up of normal conducting material the loop will undergo harmonic motion in the vertical plane which will be damped (due to resistances within the wire). The currents generating the Lorentz forces will dissipate and the loop will eventually fall down. However, let's consider the wires to be fashioned from superconducting material. In this case things change due to the Meissner effect which induces currents such that the flux $\phi$ through the loop remains constant at all times. With reference to the notation set up in Fig.~\ref{Romer}, the flux at loop position $x$ is given by $\phi(x)=w B x$. The flux due to the induced current $I(x)$ is given by the usual formula $L I$, where $L$ is the loop inductance. Thus, the Meissner effect tells us that: $L I(x)=w B x$. Using this result and noting that  the loop's potential energy is $V(x)=\frac{1}{2}L I^2(x)-mgx$,  one can see that this system corresponds to a  gravitationally driven harmonic oscillator.

\section{Model}
\begin{figure}[h!]
    \begin{center}
\includegraphics[scale=.4]{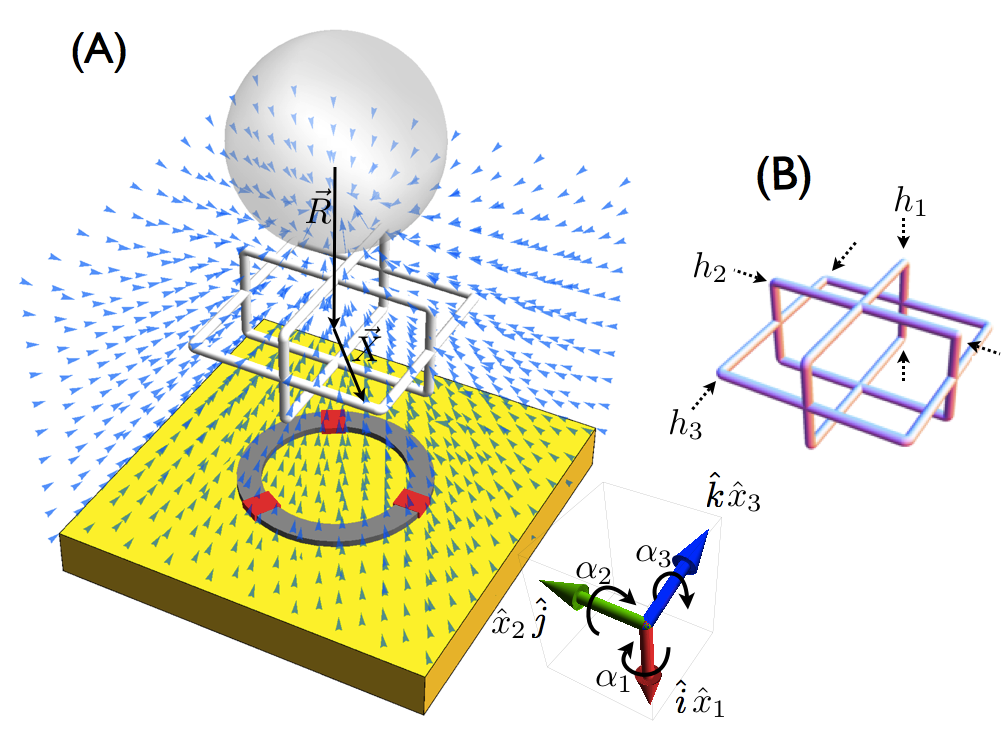}
\end{center}
\caption{Cluster of three insulated superconducting loops levitating in a magnetic field generated by a magnetized sphere with magnetization vector directed towards the positive $\hat{i}$ direction. (A) In the laboratory frame the axes are labeled $\{\hat{x}_j\}_{j=1}^3$. For each axes we have angular coordinates $\alpha_i$ (labeled accordingly) corresponding to rotation around that axes in a counterclockwise sense. The reference system has the center of the sphere as origin. The loops are considered non-intersecting (i.e. no electrical conduction between the loops).  The dimensions of the sides of the cluster of loops parallel to axis $x_i$ we take to be $h_i$. We depict the magnetic vector field generated by the spherical magnet and the nearby flux qubit (which we take to be a 3-junction phase qubit), sitting under the cluster on a yellow substrate.
(B) More detailed view of the cluster of loops.
}
\label{System}
\end{figure}

The above 2D toy model is obviously not completely trapped if naively generalised to 3D. To achieve stable three dimensional trapping we must generalise the above toy model in a more sophisticated manner. Let us now consider the system sketched in Fig.~\ref{System}. We consider a spatially strongly inhomogeneous  magnetic field generated by a fixed magnetized sphere placed above a cluster (the motional resonator) of three non-intersecting loops  oriented along mutually orthogonal planes. This particular shape will allow the potential energy of the system to have a stable minimum and also leads to a diagonal inductance matrix, so that the current flowing in one loop does not induce any currents in the other loops. A flux qubit is placed at a certain distance from the center of the cluster of loops. The inductive coupling between the flux qubit and the cluster will provide us with the control required to establish motional cooling of the levitated cluster.
In the following we first study the system classically to establish full 3D trapping, translation/rotation oscillation frequencies at the trap minimum,  before studying the quantum mechanics and motional cooling protocols to cool the vertical translational degree of freedom towards the ground state.


The total  Hamiltonian $H$ is the sum of terms involving the motional resonator ($H_r$), the qubit ($H_q$) and the interaction ($H_I$):

\begin{equation}
H=H_r+H_q+H_I\;\;,
\end{equation}
and in the following sections we develop each of these individual Hamiltonians.


\subsection{Resonator}
\label{section:Resonator}
The Hamiltonian of the cluster of loops (which we will now denote as the motional resonator or simply  resonator), can be written in the following form:

\begin{equation}
H_r=K_{\mathrm{CM}}+K_{\mathrm{rot}}+V\;\;,
\label{ResonatorHamiltonian}
\end{equation}

where $K_{\mathrm{CM}}$ is the kinetic energy due to the translational motion of the center of mass, $K_\mathrm{{rot}}$ is the rotational kinetic energy and $V$ is the effective potential energy as a function of the translational and rotational degrees of freedom.
We will now define some notation used in the following.
\begin{itemize}
\item $m$: mass of the resonator.
\item $\vec{X}$: vector position of the various part of the resonator in a co-rotating reference frame with origin at the center of mass.
\item $\vec{R}$: coordinates of the center of mass in the laboratory reference frame.
\item $\tilde{\vec{r}}$: vector position of the various part of the resonator in the laboratory frame.
\item $\vec{r}=\tilde{\vec{r}}-\vec{R}$
\end{itemize}
Note that the origin of the laboratory frame is taken to be at the center of the magnetic sphere (see Fig.~\ref{System}).\\
Because of the rigid body properties, the only allowed relation between $r(\vec{X},t)$ and $\vec{X}$ is given by $r(\vec{X},t)=O(t)\vec{X}$, where $O=e^{\sum_i\alpha_i(t) T_i}$ (with $(T_i)_{jk}=\epsilon_{ijk}$) is a rotation matrix and $\alpha_i\in[-\pi,\pi]$. We can then define the angular velocity vectors as: $\Omega_i=O(t)^{-1}\dot{O}(t)=\dot{\alpha}_iT_i$. The motion of the rigid body is thus completely determined by the set of variables $(\vec{R},\vec{\alpha})$.\\
Given these definitions, the inertia tensor of the resonator is $\mathbb{I}_{ij}=\int dV_r\rho(\vec{X})\left(\vec{X}^2\delta_{ij}-\vec{X}_i\cdot\vec{X}_j\right)$. The kinetic energies are defined with respect to the reference system in Fig.~\ref{System} as:

\begin{eqnarray}
K_\mathrm{CM}&=\frac{1}{2} m\sum_{i} \dot{R}_i^2\;\;,\\
K_{\mathrm{rot}}&=\frac{1}{2}\sum_{i,j} \mathbb{I}_{ij}\Omega_i\Omega_j\;\;,
\end{eqnarray}

The potential $V$ is just the sum of the flux energy due to the current flowing in the loops and the gravitational potential energy:

\begin{equation}\label{eq:potential_energy}
V=\frac{1}{2}\sum_{a=1}^3L_aI_a^2-m g R_1\;\;,
\end{equation}

where the index $a=1,2,3$ labels the loops accordingly to the direction perpendicular to the plane of the loop (where $1\rightarrow \hat{i}, 2\rightarrow \hat{j}, 3\rightarrow \hat{k}$), $L_a$ is the inductance of the loop $a$ and $I_a$ is the current flowing in the loop $a$. By the symmetry of the resonator the mutual inductances between the loops is zero. The currents are just functions of the six degrees of freedom (translations, rotations), describing the motion of the resonator via equations originating from the Meissner effect:

\begin{equation}\label{eq:fluxes}
\Delta\phi_a (\vec{R},\vec{\alpha})+L_a I_a(\vec{R},\vec{\alpha})=0\;\;,
\end{equation}
where $\Delta\phi_a(\vec{R},\vec{\alpha})=\phi_a(\vec{R},\vec{\alpha})-\phi_a(\vec{R}(0),\vec{\alpha}(0))$ is the difference in magnetic flux threading loop $a$ when the system is in the configuration labeled by $(\vec{R},\vec{\alpha})$ and when the system is in its initial configuration $(\vec{R}(0),\vec{\alpha}(0))$.  Any infinitesimal change in flux due to an infinitesimal displacement/rotation induces a supercurrent whose action is to restore the loop's position/orientation. The stronger this restoring force is the higher the oscillation frequency will be. We now compute the dependence of the magnetic flux on the system's configuration. 

We start by calculating the magnetic flux threading the loops. The vector potential $\vec{A}$ generated by a sphere with homogeneous magnetization vector $\vec{M}$ calculated at the point $\vec{r}+\vec{R}$ (where  $\vec{R}$ is the center of mass position vector referenced from the centre of our coordinate system - the centre of the sphere),  is just:

\begin{equation}
\vec{A}(\vec{r};\vec{M},\vec{R})=\frac{\mu_0}{4\pi}\frac{1}{|\vec{r}+\vec{R}|^3}\vec{M}\wedge(\vec{r}+\vec{R})\;\;.
\end{equation}

If we denote with $\vec{\Sigma}_a$ the area vector of loop $a$, then the flux through this loop can be expressed as:

\begin{equation}
\phi_a(\vec{R},\vec{\alpha})=\int_{\Sigma}\vec{B}\cdot d\vec{\Sigma}_a=\int_{\partial\Sigma_a}\vec{A}\cdot d \vec{r}\;\;.
\label{Flux}
\end{equation}


In a first order approximation and by supposing that the initial position of the resonator is such that $\alpha_i=0$, we have:

\begin{eqnarray}
\Delta\phi_a=\frac{\mu_0}{4\pi}(\vec{K_a}\cdot[\vec{r}-\vec{R}(0)\wedge\vec{\alpha}]+\vec{Q}_a\cdot[\vec{M}\wedge\vec{\alpha}])\;\;,
\label{deltaf}
\end{eqnarray}
where the vectors $\vec{K}_a$ and $\vec{Q}_a$ are easily calculated from the magnetic field and the sphere magnetization. By using Eq. (\ref{eq:potential_energy}) and Eq. (\ref{eq:fluxes}) we can expand the potential energy to second order in spatial and angular coordinates as:

\begin{equation}
E=\frac{1}{2}E_{ij}\zeta^i \zeta^j\;\;,
\label{eq:energySecondOrder}
\end{equation}
where $\vec{\zeta}=\{x,y,z,\alpha_x,\alpha_y,\alpha_z\}$, is a vector whose components are all the parameters describing the motion and where the matrix elements $E_{ij}$ depend on the vectors defined just above.\\

In Section \ref{Results} we will show how certain choices of the parameters of our system can lead to a 
decoupled degree of freedom (the vertical direction with associated variable $R_1$),  with a quite high  oscillating frequency $\omega_r$. This frequency is also well resolved with respect to the frequencies of the other modes (the difference between frequencies is bigger than the linewidth of the resonator), hence it can be well resolved by the qubit coupling.  Note however, that this mode is strictly decoupled from other rotational and motional modes only to second order in perturbation theory hence it is important that the system be initialized not too far from equilibrium. 
We will consider the quantized variable corresponding to small deviations from the initial position  along the $\hat{i}$ axis: $x\equiv R_1-x_0$ ($x_0$ being the $x$ component of the vector $\vec{R}_0$), whose quantum operator we define as:

\begin{equation}
\hat{x}=(\sqrt{\frac{\hbar}{2m\omega_r}})(\hat{a}+\hat{a}^\dagger)\;\;.
\label{eq:xQuant}
\end{equation}
The dynamics of this variable is governed by an harmonic Hamiltonian whose frequency $\omega_r$ can be obtained through the energy given above in Eq. (\ref{eq:energySecondOrder}).

\subsection{Qubit}

The flux qubit is considered to have characteristic frequency $\omega_q$ and to be driven by a classical field of frequency $\omega_d$ slightly detuned by $\omega_q$. The detuning parameter is $\delta=\omega_d-\omega_q$. The field drives the qubit with Rabi frequency $\Omega$. The Hamiltonian in the rotating frame with frequency $\omega_d$ is:

\begin{equation}
\hat{H}_q=-\frac{\delta}{2}\hat{\sigma}_z+\frac{\Omega}{2}\hat{\sigma}_x\;\;.
\end{equation}

\subsection{Interaction}

The classical interaction Hamiltonian for the inductive coupling between the loops of the cluster and the loop of the flux qubit is given by:

\begin{equation}
H_I=M_{rq}I_r I_q\;\;,
\end{equation}
where $M_{rq}$ is the mutual inductance between the qubit and the loop of the resonator which is horizontal in the initial position and $I_r=I_1$ and $I_q$ are the currents flowing in that loop and in the qubit. Because we are considering small deviations from the equilibrium initial position, we neglect the mutual interactions between the other loops or due to the angular motion of the object. We will consider the dependence of the coupling in the vertical direction by performing a first order expansion.  We  expand $I_r$ to first order  in small deviations from the initial position using (\ref{deltaf}). In the quantized version we also replace $I_q$ with $I_q \hat{\sigma}_z$ (see for example \cite{Wallquist}). Denoting $D_I(0)=\frac{\partial I_r}{\partial x}{\Big|_{\vec{R}(0)}}$ as the derivative of the current evaluated at the initial cluster position and remembering (\ref{eq:xQuant}) we write the quantized interaction Hamiltonian as:

\begin{equation}
\hat{H}_{I}=\hbar~\frac{\lambda}{2}(\hat{a}+\hat{a}^\dagger)\hat{ \sigma}_z\;\;,
\end{equation}

with:

\begin{equation}
\lambda=\left(\sqrt{\frac{2}{m\hbar\omega_r}}\right)MD_{I_l}(0) I_q \;\;.
\end{equation}

The mutual inductance is simply calculated as:
\begin{equation}
M=\frac{\mu_0}{4\pi}\int\frac{d\vec{s}d\vec{s}^{~\prime}}{R}\;\;,
\end{equation}
where $d\vec{s}$ and $d\vec{s}^\prime$  are vectors tangent to the two loops  and $R$ is the distance between the infinitesimal part of each loop and the integration is along each of the two loops.\\

The final quantized expression for the Hamiltonian describing the motion of the cluster in the vertical direction  and its interaction with the qubit is (setting $\hbar=1$):

\begin{equation}
\hat{H}=-\frac{\delta}{2}\hat{\sigma}_z+\frac{\Omega}{2}\hat{\sigma}_x + \omega_r \hat{a}^\dagger \hat{a} + \frac{\lambda}{2}(\hat{a}+\hat{a}^\dagger)\hat{\sigma}_z\;\;.
\label{HT}
\end{equation}
In order to be in the motional quantum regime we must devise cooling schemes to bring the resonator close to its motional ground state as described in the next section.

\section{Cooling}

\subsection{Cooling scheme A: Introduction}

\begin{figure}[h!]
    \begin{center}
    \includegraphics[scale=0.6]{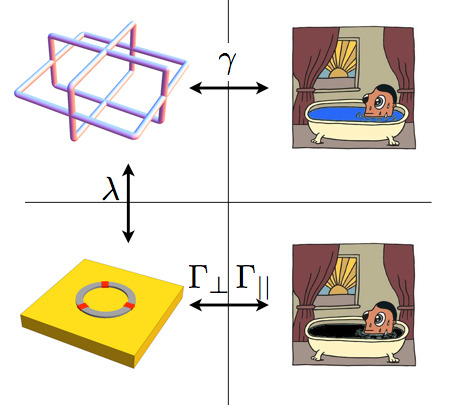}
    \end{center}
\caption{Motional Cooling Scheme: we inductively couple (with strength $\lambda$), the motion of the levitated resonator with a nearby flux qubit. The resonator looses energy at rate $\gamma$ due to coupling to a local phonon bath while the qubit experiences both decay and dephasing at rates $\Gamma_\bot$ and $\Gamma_{||}$. In addition we drive transitions in the qubit at rate $\Omega$.}
\label{CoolingScheme}
\end{figure}

Following \cite{Rabl} we will consider  the resonator and the qubit coupled to separate thermal baths and interacting with each through the coupling $H_I$ (see Fig.~\ref{CoolingScheme}).

The system's quantum motional state, $\hat{\rho}$ evolves under the following master equation: 

\begin{equation}
\dot{\hat{\rho}}=-i[\hat{H},\hat{\rho}]+\hat{\mathcal{L}}_{\Gamma}(\hat{\rho})+\hat{\mathcal{L}}_{\gamma}(\hat{\rho})\;\;.
\label{master1}
\end{equation}

The qubit Liouville operator contains decay and dephasing with rates $\Gamma_\bot$ and $\Gamma_{||}$: 

\begin{eqnarray}
\hat{\mathcal{L}}_{\Gamma}(\hat{\rho})=&\frac{\Gamma_\bot}{2}(N_{q}+1)(2\hat{\sigma}_{-}\hat{\rho}\hat{\sigma}_{+}-\hat{\sigma}_{+}\hat{\sigma}_{-}\hat{\rho}-\hat{\rho}\hat{\sigma}_{+}\hat{\sigma}_{-})+\nonumber\\
&+\frac{\Gamma_\bot}{2}(N_{q})(2\hat{\sigma}_{+}\hat{\rho}\hat{\sigma}_{-}-\hat{\sigma}_{-}\hat{\sigma}_{+}\hat{\rho}-\hat{\rho}\hat{\sigma}_{-}\hat{\sigma}_{+})+\nonumber\\
&+\frac{\Gamma_{||}}{2}(\hat{\sigma}_z\hat{\rho}\hat{\sigma}_z-\hat{\rho})\;\;,
\end{eqnarray}

where $N_q=(e^{\frac{\hbar\omega_q}{k_B T}}-1)^{-1}$, and where $T$ is the temperature of the qubit bath.
The Lioville operator for the motional resonator is:
\begin{eqnarray}
\hat{\mathcal{L}}_{\gamma}(\hat{\rho})=&\frac{\gamma}{2}(N_{th}+1)(2 \hat{a}\hat{\rho} \hat{a}^{\dagger}-\hat{a}^{\dagger}\hat{a}\hat{\rho}-\hat{\rho} \hat{a}^{\dagger}a)+\nonumber\\
&+\frac{\gamma}{2}(N_{th})(2 \hat{a}^{\dagger}\hat{\rho} \hat{a}-\hat{a} \hat{a}^{\dagger}\hat{\rho}-\hat{\rho} \hat{a} \hat{a}^\dagger)\;\;,
\end{eqnarray}
where the mechanical dissipation factor $\gamma$ has been introduced, and the equilibrium phonon occupation number is $N_{th}=(e^{\frac{\hbar\omega_r}{k_B T}}-1)^{-1}$.\\
We now trace out the qubit to find an effective equation of motion for the resonator. From this we will see that under certain cases we can obtain a motional  cooling process bringing the cluster towards its motional ground state. We will consider the two different temperature regimes studied in \cite{Rabl}.

\subsubsection{Low Temperature Regime}

Here we consider a regime where $\lambda\sqrt{N_{th}+1/2}\ll\Gamma_{\bot},\omega_r$, which is equivalent to the Lamb-Dicke regime in laser cooling (see \cite{Rabl} and references within). The first condition tells us that 
the qubit damps much faster than energy can be transferred to it from the resonator. By observing that the parameter $\lambda$ describes the position displacement kick of the resonator when the qubit undergoes a spin flip (see \cite{Rabl}), the second approximation essentially says that the back-reaction on the resonator following a qubit flip is small relative to the oscillator energy. \\

Thanks to these assumptions one can write the full density matrix as $\hat{\rho}(t)\simeq \hat{\rho}^0_q\otimes\hat{\rho}_r(t)$ where $\hat{\rho}_q^0$ is the qubit steady state density matrix. In this way one can obtain an effective master equation for the resonator after having traced out the qubit. The result can be encoded in a new effective mechanical damping factor given by $\Gamma=\Gamma_C+\gamma$ with $\Gamma_C=S(\omega_r)-S(-\omega_r)$ where $S(\omega)$ denotes the qubit fluctuation spectrum:

\begin{equation}
S(\omega)=\frac{\lambda^2}{2}\mathrm{Re}\int_0^\infty \, e^{i\omega\tau}\,d\tau\, \left\{\langle\hat{\sigma}_z(\tau)\hat{\sigma}_z(0)\rangle_0-\langle\hat{\sigma}_z\rangle_0^2\right\}\;\;,
\end{equation}

and where $\langle \cdot \rangle_0$ denotes the steady state expectation \cite{Rabl}. What interests us is the final motional occupation number which is given by:

\begin{equation}
n_{LD}=\frac{\gamma N_{th}}{\Gamma_C}+N_0\;\;,
\label{nf}
\end{equation}

where $N_0=S(-\omega_r)/\Gamma_C$, and the subscript stresses the Lamb-Dicke regime we are supposing.

\subsubsection{High Temperature Regime}

As one can see from (\ref{nf}), for low enough $N_{th}$ the final occupation number is given by $N_0$. Increasing the temperature of the motional bath we see that the final occupation number grows linearly with $N_{th}$. However, when $N_{th}$ is so large the previous  assumptions $\lambda\sqrt{N_{th}+1/2}\ll\Gamma_{\bot},\omega_r$ do not hold true anymore it turns out \cite{Rabl} that, under the temperature independent assumptions ($\lambda\ll\Gamma_{\bot},\omega_r$), one can study situations in which the final occupation number is given by the more general formula:




\begin{equation}
n_f=N_{th}\left( \zeta+\frac{1-\zeta}{1+\zeta e^{I_1/(N_{th}\zeta\eta^2)}}\right)\;\;,
\label{Occupation}
\end{equation}
with parameters defined in \ref{appendix:high_temperature}. As described in that section, in order to obtain this formula one must consider  the resonator`s initial quantum state is given by the coherent state $|\alpha\rangle\langle\alpha|$ with $|\alpha|^2\sim N_{th}$ a coherent state labeled by the parameter $\alpha$ as initial condition. Given this assumption the ``renormalized'' cooling rate $\Gamma_C$ is an $\alpha$ dependent quantity. 

\subsection{Cooling scheme B}

Rather than use the scheme depicted in Fig \ref{CoolingScheme}, one can also cool the motion of the resonator (at least in the case of near resonance with the qubit) by  performing a sequence of repeated measurements on the flux qubit inductively coupled to the resonator's position  ~\cite{Li}.\\
Using a flux qubit made out of three superconducting loops with four Josephson junctions one \cite{Paauw,Fedorov} can tune the magnetic energy bias $\hbar \delta$ to zero. We also assume that the inductive coupling strength is much smaller than the qubit frequency ($\lambda\ll\omega_r$). This will allow us to perform a rotating wave approximation on the Hamiltonian of the full system. We consider the system to be initially prepared in the separable state $\rho_0=\ket{g}\bra{g}\otimes\rho_r$ with the resonator in the thermal state $\rho^n_r=\bra{n}\rho_r\ket{n}=\frac{e^{-n\beta\hbar\omega_r}}{1-e^{-\beta\hbar\omega_r}}$. Then, one performs $N$ measurements on the qubit  (with associated operator given by $\ket{g}\bra{g}$ and  $\ket{e}\bra{e}$  where $g$ (e) is the ground (excited) state) at times $t_j=j\tau$ where $\tau$ is  short time interval. One finds that, provided the measurement outcome is always $\ket{g}\bra{g}$, at the conclusion of these $N$ measurements the state of the resonator is \cite{Li}: 

\begin{equation}
\rho_r(N)=\sum_{n\ge 0}\frac{|\mu_n|^{2N}\rho^n_r\ket{n}\bra{n}}{P(N)}\;\;,
\end{equation}

where $\mu_0=e^{i\Delta\tau/2}$, $\mu_n=e^{-i(n-1/2)\omega_r\tau}(\cos{\Omega_n}\tau+i\sin{\Omega_n\tau}\cos{2\theta_n})$, $\Omega_n=\sqrt{(\Delta-\omega_r)^2/4+g^2 n}$ ($n\ge 1$), $\tan{2\theta_n}=2 g\sqrt{n}/(\Delta-\omega_n)$ ($n\ge 1$) and where $N$ indicates the number of measurements. $P(N)$ is the probability this sequence of measurement outcomes occurs.

\section{Results}
\label{Results}

Here we consider our 3D model in a more realistic setting and adopt physical parameters in the table shown in section \ref{Table}. Using this physically realistic set of parameters we now examine in more concrete detail the classical and quantum mechanics of the resonator and the levels of cooling achieved by the above outlined methods.\\

\subsection{Classical Motion}
\begin{figure}[h!]
    \begin{center}
\includegraphics[scale=0.8]{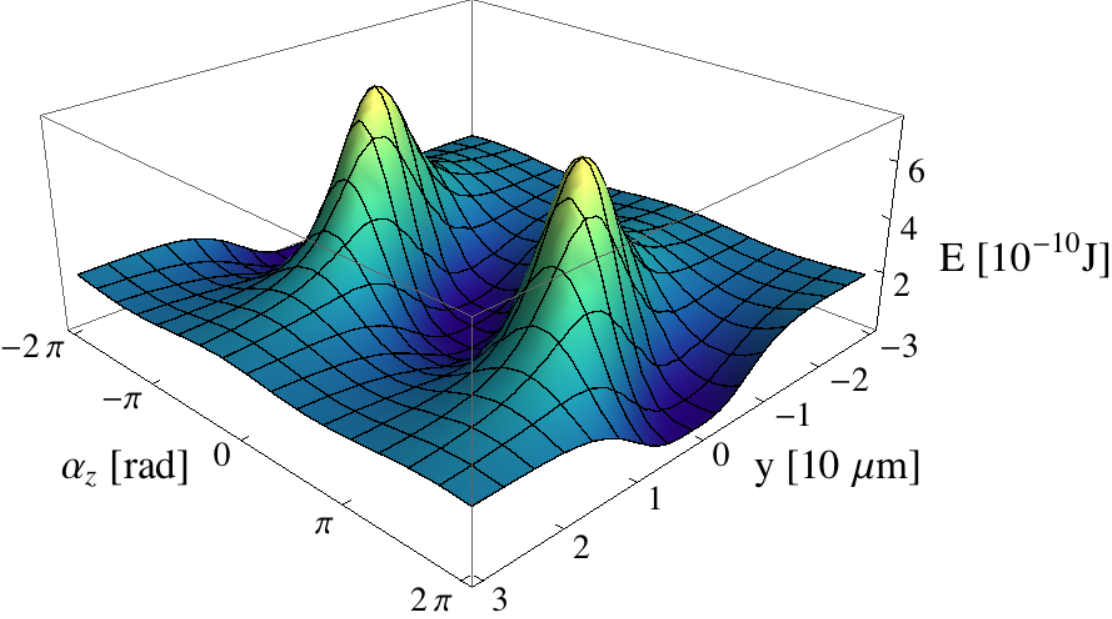}
\end{center}
\caption{Exact numerical evaluation of the potential energy as a function of the horizontal translational degree of freedom $y$ (with origin the stable point) and the angular degree of freedom $\alpha_z$.  Because the potential is confining to third order in these degrees of freedom the overall system is stable. }
\label{EnergyYazBlue}
\end{figure}

We now examine the classical motion of the resonator by focusing in particular on its stability and its frequency of  oscillation.\\ 
The effect of the gravity in the potential energy $V(x)$ causes a small and negligible shift of the potential minimum.\\
By considering the magnetization of the sphere to be aligned along the vertical $x$ direction and expanding to second order the potential energy around its minimum, one finds that the $x$-motion decouples from the other directions. One finds further the existence of three degrees of freedom which are ``zero modes'', i.e. do not contribute to the energy up to third order. One zero-mode is related to the rotational symmetry around the vertical axes. The other two are due to the coupling between  the variables $y$($z$) and $\alpha_z$($\alpha_y$).  however, if one considers these contributions to  higher orders these degrees of freedom contribute to the energy value to give overall stability  as shown in  Fig.~\ref{EnergyYazBlue}. 
For a cluster of loops of ''typical ``dimensions (as defined below) of  $(1,10,10)~\upmu$m and with a wire thickness of $0.1~\upmu$m  the root mean square thermal motion of the cluster at $15$ mK is $\sim 10^{-12}$ m along the spatial directions and $10^{-6}$ radians along the $\alpha_y$ and $\alpha_z$ angular directions (while there is symmetry along the other angular direction). \\

\begin{figure}[h!]
    \begin{center}
\includegraphics[scale=0.2]{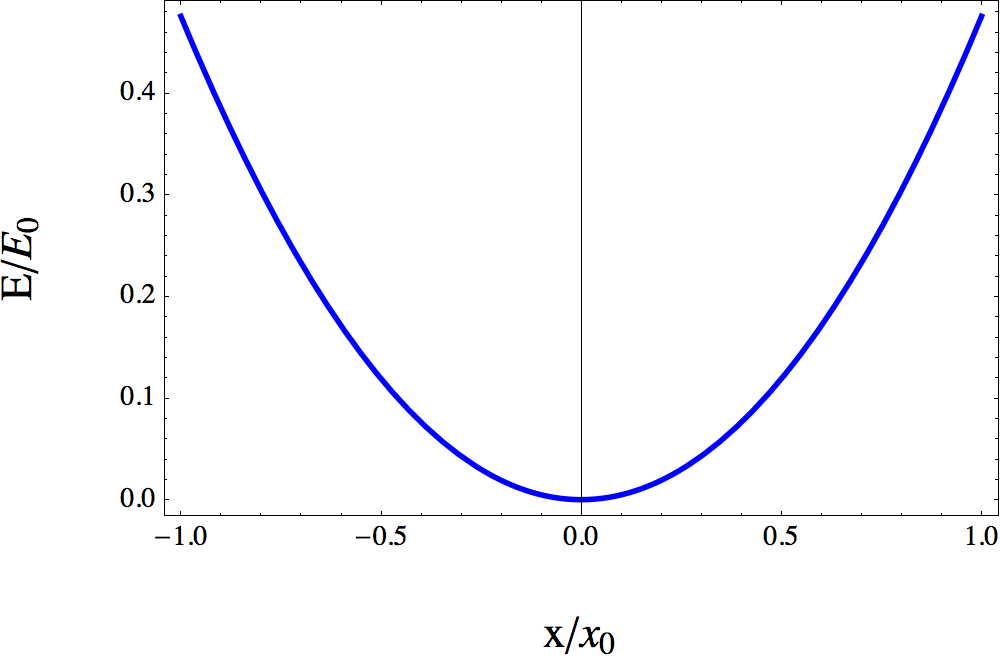}
\end{center}
\caption{Exact numerical evaluation of the potential energy (in units of $E_0=\hbar\omega_r$) as a function of the vertical translational degree of freedom $x$ (with origin the stable point and in unit of $x_0=\sqrt{\frac{\hbar}{m\omega_r}}$).}
\label{EnergyX}
\end{figure}

\begin{figure}[h!]
    \begin{center}
\includegraphics[scale=0.13]{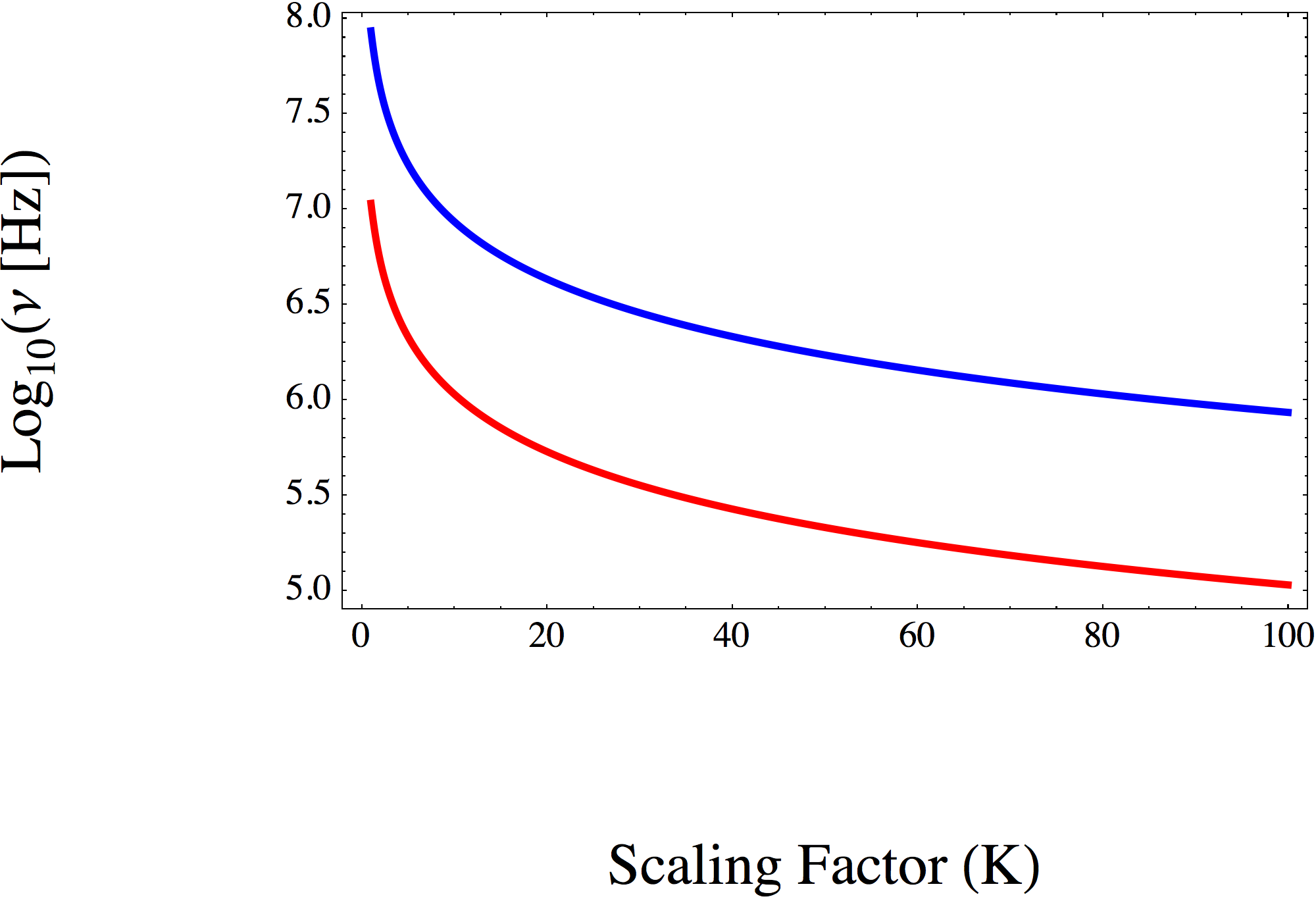}
\end{center}
\caption{Translational mode frequency as a function of system size.  The frequency along the $\hat{x}$ direction (blue line) and $\hat{y}$ (and for symmetry reasons $\hat{z}$) direction (red line) are plotted as a function of the scaling factor $k$ which defines the dimension of the system. The radius of the magnetized sphere is defined as $R_{sp}=k~ \upmu$m. The dimension of the parallelepiped enveloping the cluster of loops is $k\times(0.1,1,1)\upmu$m and the thickness of the wires is $k~\times 0.01~\upmu$m.  The cluster mass scales as $k^3$. When considering different values for the scale factor, the initial position of the center of the cluster of loops is chosen to keep the distance between the center of the sphere to the top of the cluster fixed at $\sim 1.1 ~R_{sp}$ where $R_{sp}$ is the radius of the sphere.
The frequencies are higher for smaller sizes of the system. As  ``typical'' parameters for the system we will consider the ones given by a scaling factor $k=10$. }
\label{FreqFact}
\end{figure}

The frequency of the translational motions  depends on the physical dimensions of the cluster of loops. In Fig.~\ref{FreqFact} the frequency in the $x$ and $y$-direction is plotted as a function of a scaling factor $k$ for the system.
The values for all the parameters are explicitly shown in the table in section \ref{Table}.\\
The frequencies decrease as the scaling grows, showing that the trapping in the translational degrees of freedom is becoming less tight, but still present for a large range of size scales. In regards of the rotational degrees of freedom, even when the system is scaled up with a factor 10 with respect to the typical parameters, numerically we find that the trapping is still achieved in the small oscillation regime (see section \ref{section:Resonator}).\\

In summary, using the parameters shown the table in section \ref{Table}, for small loops and very high magnetic field inhomogeneities the cluster of loops experiences stable, levitated magnetic trapping. The frequency of oscillation along the $x$-direction is larger than in the other directions making it easier to cool. For these reasons, in the following we quantize the $x$-translational motion.

\subsection{Superconductivity and Dissipation}

\begin{figure}[h!]
    \begin{center}
\includegraphics[scale=0.7]{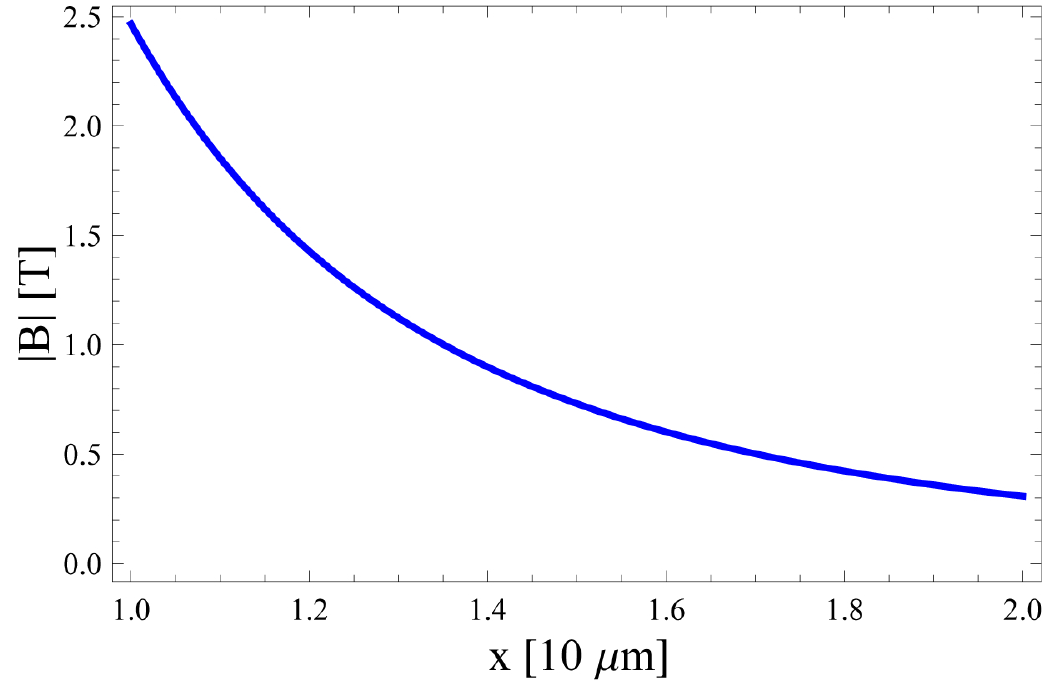}
\end{center}
\caption{Magnetic field intensity along the vertical $\hat{x}$.}
\label{BField}
\end{figure}

The superconducting loops are considered as made out of Niobium-Tin (NbSn) which, in normal conditions, has very high critical magnetic field strength ($\sim 30$ T ), although, in our case, the situation is more critical due to the small thickness of the superconducting wires.\\
We now examine two types of motional decay mechanisms and estimate the resulting motional Q for the resonator`s motion. We first consider the loss in energy from the moving resonator due to its action as a dipole emitter of radiation.\\
In Fig.~\ref{BField} the intensity of the magnetic field (in Tesla) is plotted as a function of distance from the center of the magnetized sphere (radius $10\upmu$ m).

As we saw above, the motion of the cluster of loops is trapped in all three directions. Due to the Messnier effect, the greater the spatial excursion from its equilibrium position the greater the current density inside the wires. Considering the physical parameters given in the table in section \ref{Table} we find typical current  densities of $\sim(10^8,10^8,10^7)$A/m${}^2$ (respectively in the loops perpendicular to the $\hat{x}$, $\hat{y}$, $\hat{z}$ directions).Considering each current to be described by $I=I_0e^{-i\omega t}$, where $\omega$ is the top frequency in each direction, by approximating the loops to be circular we can calculate the power dissipated by the oscillating currents as electromagnetic radiation as (see for example \cite{McDonald}):

\begin{equation}
P_{\mathrm{rad}}=\frac{1}{2}\frac{\pi}{6}\frac{4\pi}{c}\frac{2\pi r}{\lambda}\frac{I_0}{2}\;\;,
\end{equation}

where $r$ is the radius of the loop and $\lambda=c/\nu$ (where $c$ is the speed of light). From this we observe that the power radiated is absolutely negligible. \\
However, the main source of dissipation is the  viscous drag of flux lines oscillating 
inside the pinning wells inside the superconducting wires  \cite{Schilling}. One can estimate the motional quality factor $Q=\frac{\omega_r}{\gamma}$ for a system similar to (the two dimensional version of) the one presented here to be $Q\approx10^{11}$. In our calculations we will consider $Q\approx 10^{10}$. Such an enormously high motional quality factor is already known in levitating systems ~\cite{Zoller} ($Q \sim 10^{12}$), although typically one finds motional $Q$ values in the range $\sim 10^3\slash 10^6$ for cavity optomechanical experiments~\cite{Kippenberg}. The prospect for ultra-large motional $Q$ is one of the primary benefits of our scheme.

\subsection{Coupling between the resonator and the flux qubit.}
\begin{figure}[h!]
    \begin{center}
\includegraphics[scale=0.3]{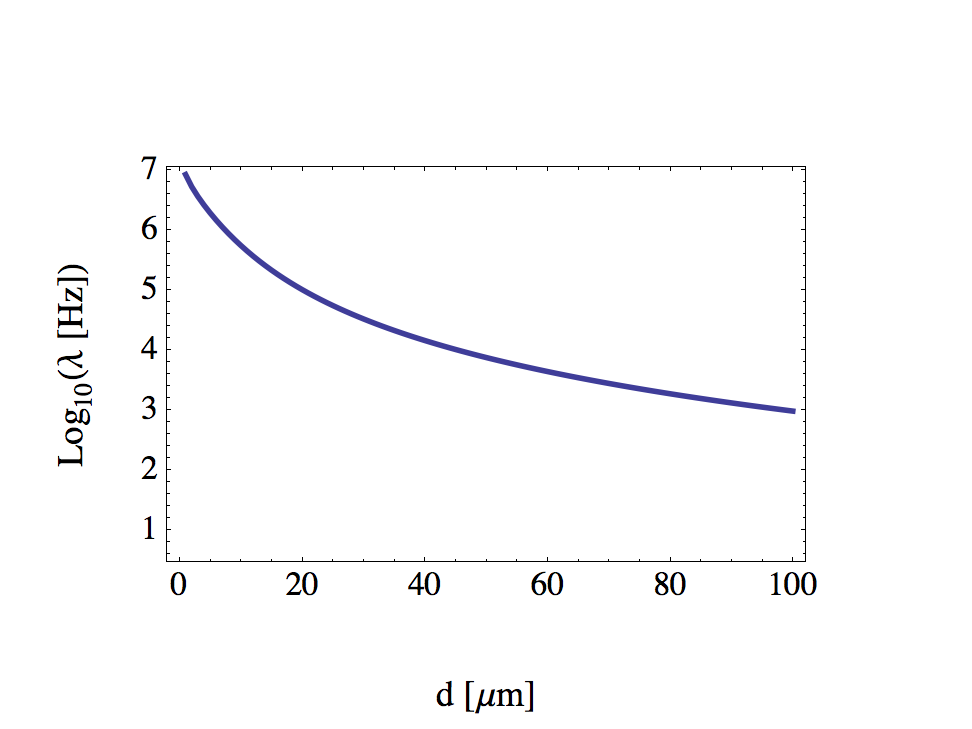}
\end{center}
\caption{Coupling constant between the qubit and the resonator. The coupling arise from the mutual inductance between the loops of the resonator and the qubit loop. By supposing the loop of the qubit to be placed in a plance orthogonal to the vertical direction, the contribution to the coupling comes, in a good approximation, from the loop of the resonator perpendicular to the $x$ direction. The distance $d$ is defined as the distance between the center of mass of the resonator and the center of the qubit loop. The qubit loop is supposed to have a $5\upmu$m long radius. }
\label{Coupling}
\end{figure}

In this section we will study the coupling between the resonator and the flux qubit. The possibility to tune this coupling arises from  the (distance dependent) inductive nature of the coupling.\\
Fig.~\ref{Coupling} shows the value of the coupling strength between the qubit and the resonator as a function of the distance $d$ between the center of mass of the resonator and the center of the loop of the flux qubit the latter taken to be a circle of radius $5\upmu$m.

We observe that the coupling strength  can be adjusted over quite a large range by fixing the distance between the qubit and the resonator.

\subsection{Cooling of the motional resonator.}
\subsubsection{Cooling scheme A.}
\begin{figure}[h!]
    \begin{center}
\includegraphics[scale=.25]{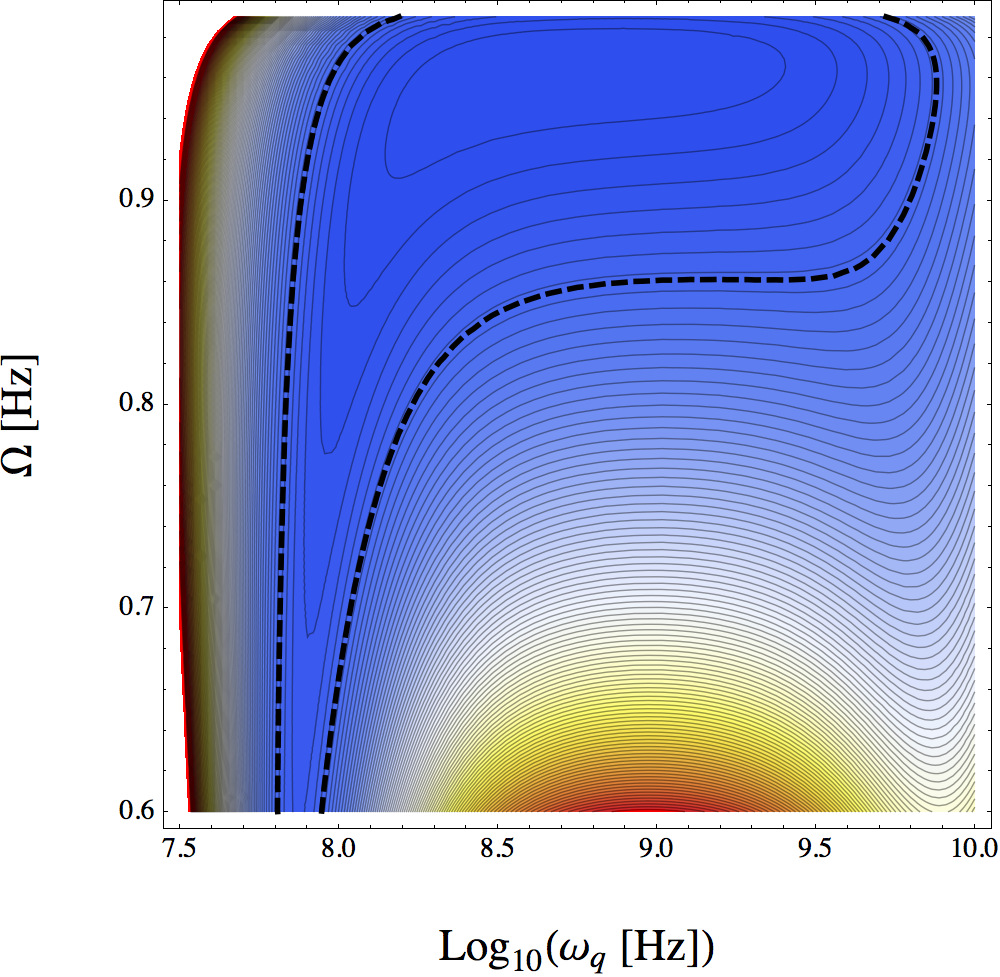}
\end{center}
\caption{Contour plot  representing the final occupation number (blue corrsponding to zero phonon and red corresponding to 200 phonons) versus the qubit parameters $\Omega$ and $\omega_q$ with $\delta$ fixed by the constraint $\delta=\sqrt{\omega_q^2-\Omega^2}$. The dashed line represents the phonon equilibrium number if the system is in equilibrium with a thermal bath with temperature $15$ mK. Cooling takes place for parameters inside the dashed line. The frequency of the resonator has been taken to be such that $\omega_r=2\pi10^7$. The coupling between resonator and qubit has been set to $10^4$. With this value, the low temperature approximation is a good one for qubit frequencies such that $\omega_q\ge 10^9$.}
\label{LowT}
\end{figure}

\begin{figure}[h!]
    \begin{center}
\includegraphics[scale=0.25]{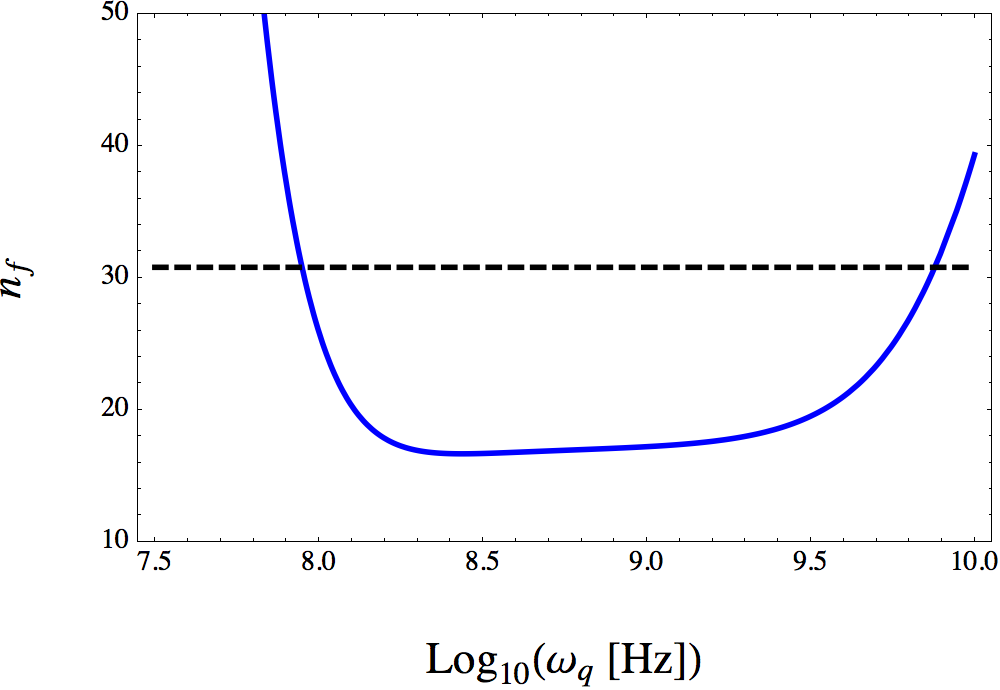}
\end{center}
\caption{Plot of the final occupation number as a function of $\omega_q$ with qubit parameters fixed as follows: $\Omega=0.95 \omega_q$ and $\delta=\sqrt{\omega_q^2-\Omega^2}$. This graph is just the cross section of fig. \ref{LowT} obtained by fixing $\Omega$. The dashed line is the same as in fig. \ref{LowT} and represents the phonon equilibrium number if the system is in equilibrium with a thermal bath with temperature $15$ mK. The frequency of the resonator has been taken to be such that $\omega_r=2\pi10^7$. The coupling between resonator and qubit has been set to $10^4$.  With this value, the low temperature approximation is a good one for qubit frequencies such that $\omega_q\ge 10^9$.}
\label{LowT2D}
\end{figure}

In this cooling scheme the motional energy of the resonator is damped away thanks to the faster decay of the flux qubit.\\
We will start by studying the low temperature limit. This means that the $N_{th}$ of the bath connected to the resonator must be such that  $\lambda\sqrt{N_{th}+\frac{1}{2}}\ll\Gamma_\bot,\omega_r$ is fulfilled.\\
In Fig.~\ref{LowT}  we plot the final motional average Fock number as a function of the qubit resonance frequency $\omega_q$ and Rabi frequency $\Omega$, while in Fig. \ref{LowT2D} we plot the same final occupation number as a function of $\omega_q$ after fixing $\Omega$ to an optimal value. In particular, we chose the Rabi frequency $\Omega$ and the detuning $\delta$  to fulfill the relation $\omega_q=\sqrt{\Omega^2+\delta^2}$. For comparison, the dashed line in Fig.~\ref{LowT} shows the equilibrium phonon number if the resonator is in equilibrium with a $15$ mK thermal bath. The cooling happens in the region defined by the dashed line. Fig.~\ref{LowT2D}, which is  a section of Fig.~\ref{LowT} obtained by fixing $\Omega$. 

\begin{figure}[h!]
    \begin{center}
\includegraphics[scale=0.25]{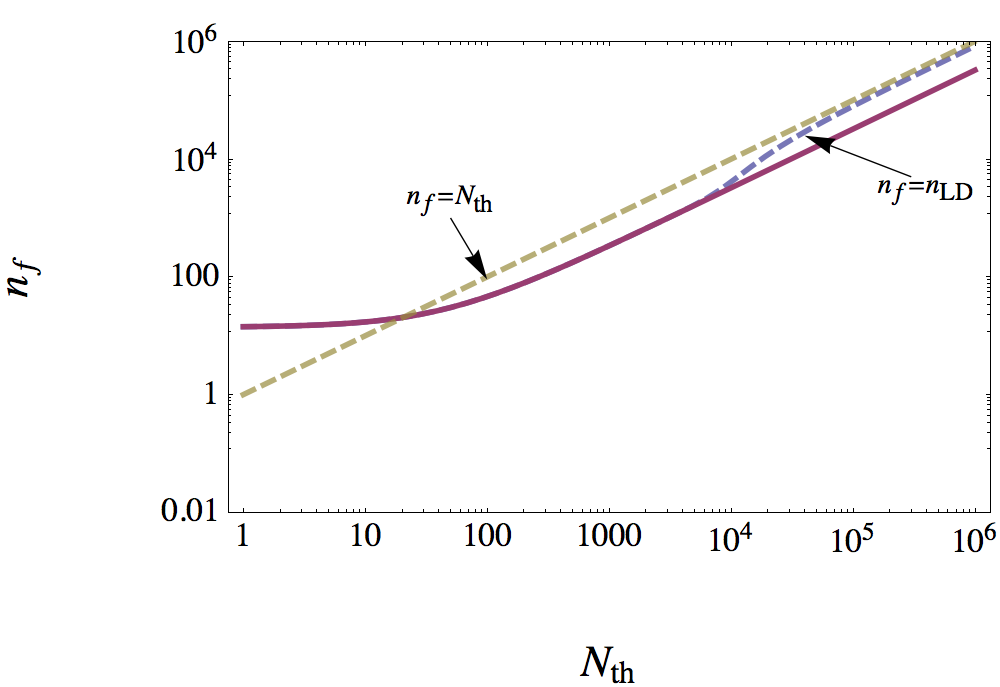}
\end{center}
\caption{Plot of the final occupation number as a function of the initial occupation number. The dashed line labeled as $n_{LD}$ represents the low temperature limit. The other dashed line represents the identity line. For this plot $\omega_r=2\pi 10^7$, $\omega_q=10^9$, $\lambda=10^5$.}
\label{HighT1}
\end{figure}

\begin{figure}[h!]
    \begin{center}
\includegraphics[scale=0.25]{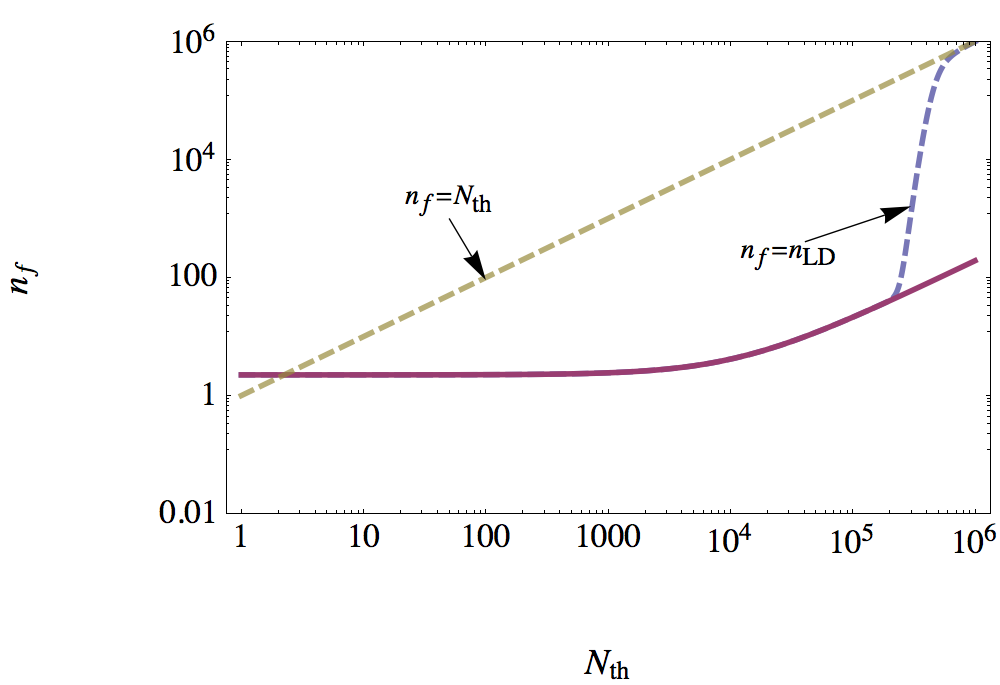}
\end{center}
\caption{Plot of the final occupation number as a function of the initial occupation number. The dashed line labeled as $n_{LD}$ represents the low temperature limit. The other dashed line represents the identity line. Here $\omega_r=10^9$, $\omega_q=10^9$, and $\lambda=10^5$. The result is that cooling is possible in this resonance situation even if the bath connected to the resonator is very hot ($T\sim100$ K).}
\label{HighT2}
\end{figure}

Now we move to a regime where no constraints on the temperature of the bath connected to the resonator are imposed so that the only relations to be fulfilled are $\lambda\ll\Gamma_\bot,\omega_r$. This will allow us to use an higher coupling constant with respect to the low (but still non-zero) temperature limit, previously considered.\\
Fig.~\ref{HighT1} shows the final expected cooled final motional Fock number $n_r$ in the case where there exist a large frequency mismatch between the qubit and the resonator. As expected the cooling is quite poor. For comparison, we also plot the dashed line labeled as $n_{LD}$ obtained from the low temperature theory extrapolated to high temperature (as given by Eq. (\ref{nf})). One can see that in the low temperature regime the two theories agree while at high temperatures they disagree. Above a certain bath temperature cooling is no longer possible. Below this bath temperature cooling is possible though in the far off-resonant case the degree of cooling is poor. \\
In Fig~\ref{HighT1}  we consider the case where the qubit and the resonator are vastly off resonance and this gives very poor cooling performance. One can expect the cooling to be far better when they are both on resonance. To achieve this, one must either reduce the qubit frequency or increase the resonator frequency, possibly via new flux qubit designs \cite{Gustavsson} or by replacing the magnetized sphere with a magnetic tip \cite{Mamin} with much larger magnetic field gradients.\\
In Fig.~\ref{HighT2} we consider the same cooling process as in Fig.~\ref{HighT2}  but with the resonator now resonant with the qubit. In this case the system shows a significant enhancement in the cooling rate which scales from $\gamma=0.1$ (off-resonance) to $\gamma+\Gamma_C\in(5\times 10^2,10^2)$ (resonant) for the regime where the cooling is still possible. The net result is that, even if the resonantor is connected to a very hot thermal bath ($N_{th}\sim 10^4$), one can cool the system to a final average phonon number around $n_f\sim 1$.


We showed that the interaction between the qubit and the resonator leads to an increase of the cooling rate. Cooling is possible even when the initial effective temperature is quite high. Cooling efficiency increases significantly when the resonator and the qubit are in resonance.

\subsubsection{Cooling scheme B.}
\begin{figure}[h!]
    \begin{center}
\includegraphics[scale=0.3]{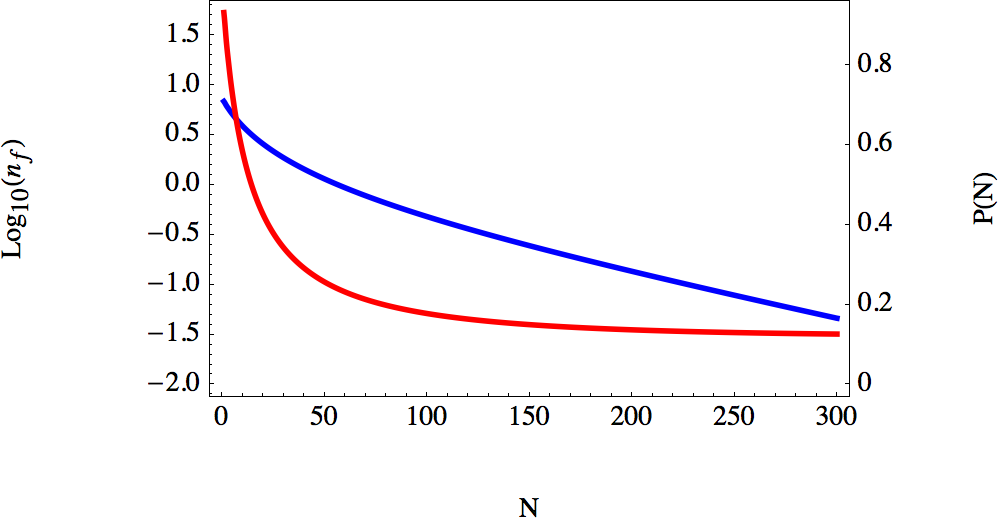}
\end{center}
\caption{Plot about the cooling by mean of measurements on the qubit. The blue line represents the plot of the final occupation number $n_f$ as a function of the number of measurements on the qubit.  The red line represents the survival probabilty $P(N)$ which, for $N\rightarrow\infty$, goes to the initial probability to find the system in the ground state. For this plot the resonator is interacting with a $T=15$mK thermal bath and we consider a resonant situation with $\omega_r=\omega_q=2.5\times10^8$ Hz, $\lambda=5\times 10^6$ Hz and $\tau=\frac{10}{\omega_r}$ s.}
\label{AltroCooling}
\end{figure}

This scheme cools down the motion of the qubit using repetited measurements on the qubit inductively coupled to the resonator. The cooling is achieved in the case in which the outcomes of the measurements always tell us that the qubit is in the ground state. This ``lucky'' sequence happens with probability that decays exponentially with the number of measurement but which, in the limit of infinite measurement does not vanish, but rather converges to the initial ground state probability.\\ 
Henceforth we suppose the qubit and the resonator are in resonance. The value of the frequency has been fixed so to maximize the results given a maximum value for the coupling $\lambda=5$ MHz. Fig.~\ref{AltroCooling} shows the final occupation number as a function of the number of measurements on the qubit. In this way it is possible to reach the ground state after a quite limited number of measurements. As explained in \cite{Li}, requiring equal times between  measurements (which would be a technical difficulty) is not important and can be relaxed by introducing random time deviations. The situation would even improve in this case. 

\section{Conclusions}

We presented here a mechanical oscillating system that is inherently non-dissipative that utilizes superconducting material levitated in a vacuum via the Meissner effect. We showed that the cluster of loops is trapped in a potential well given by a magnetic field generated by a nearby magnetized sphere. Among the total of 6 degrees of freedom of the resonator we quantized the variable associated with the vertical direction as, in a second order approximation, it is decoupled from the other degrees of freedom and it possess the largest oscillating frequency. We also showed how to cool down the motion of the system in the low temperature regime as well as in the high temperature regime through coupling to a qubit. The system allowed us to consider two different cooling schemes which work very well in the case  resonator and qubit are resonant. The system can be improved by considering stronger or higher gradient magnetic fields, which could be achieved by replacing the magnetized sphere with other geometric magnetized objects. This is not the only direction for further studies. In fact a complete analysis would quantize all the degrees of freedom and would find a way to cool all of them (for example by introducing other inductive couplings with other flux qubit).

\section{Acknowledgements}
We thank Gerald Milburn for many helpful comments and discussions.  This research was supported by the ARC via the Centre of Excellence in Engineered Quantum Systems (EQuS), project number CE110001013.


\newpage

\section{Parameter Table}
\begin{table}[ht] 
\centering      
\begin{tabular}{|c |c ||c|}  
\hline
Variable & Reference Value & Meaning\\ [0.5ex] 
\hline                    
$R$ &$10~\upmu$m & Radius of the Sphere   \\   
$R_0$ & $110\%~R + \frac{1}{2}h_1$  & Initial distance from center of sphere \\ 
$M$ & $\frac{3.7}{\upmu_0} ~{\mathrm m}^{-3}$  & Magnetization density\\ 
$\hat{M}$ & $\hat{x}$  & Magnetization vector\\ 
$\rho$ & $6.6\times 10^3$~kg~m$^{-3}$ & Superconductor mass density (close to N${\rm b_3}$Sn density)\\
$r$ & $50$~nm & Wires radius\\ 
$h_1$ & $1\upmu$m & Wires length along $x$\\ 
$h_2$ & $10\upmu$m & Wires length along $y$\\ 
$h_3$ & $10\upmu$m & Wires length along $z$\\ 
$d$ & $>1\upmu$m & Distance between qubit and cluster\\ 
$\Omega$ & $(0.3-10)$~GHz & Driving Field Frequency\\ 
$\delta$& $-(1-100)\%\Omega$ & Dephasing\\ 
$T_1$& $0.5\upmu$s &Qubit Decay Time\\ 
$T_2$& 1$\upmu$s & Qubit Dephasing TIme\\ 
$\Gamma_{\bot}$& $\frac{1}{T_1}\frac{1}{2 N_{\mathrm{TLS}+1}}$& Qubit Decay Rate \\ 
$\Gamma_{||}$& $T_2^{-1}-\frac{1}{2}T_1^{-1}$ & Qubit Dephasing Time\\ 
$\lambda$& $(10^4,10^5)$Hz & Interaction between qubit and cluster\\ 
$T$& $15$~mK & Temperature\\ 
$Q$& $10^{10}$ & Estimated Quality Factor\\ 
$\omega_r$& $2\pi 10$~MHz & Mechanical Angular Velocity\\ 
$\gamma$& $\frac{\omega_r}{Q}$ & Mechanical Energy-Damping Rate\\ 
$N_{\mathrm{q,r}}$& $(e^{\frac{\hbar \omega_{\mathrm{q,r}}}{k_B T}}-1)^{-1}$ &Thermal Equilibrium Occupation Number\\ 
$I^\prime_q$&$10~\mathrm{nA}$&Qubit current\\
$m$&$4 \times10^{-15}~\mathrm{kg}$&Cluster of loop mass\\
$D_{I_z}(0)$&$\approx 7\times10^5 ~\mathrm{A/m}$&Current derivative calculated at Origin\\[1ex]       
\hline     
\end{tabular} 
\label{Table}  
\end{table} 

\appendix
\section{High temperature regime final occupation number}\label{appendix:high_temperature}

By supposing the initial state of the resonator to be a coherent state parametrized by the complex variable $\alpha$, and by supposing the time scale of the interaction to be much longer than the time scale with which the resonator interacts with its bath ($\lambda T_1\ll 1$) one can obtain, from the master equation of the system, the bloch equation given by:

\begin{equation}
<\dot{\vec{S}}>=\mathbf{A}<\vec{S}>-i\lambda(e^{-i\omega_r t}\alpha+e^{i\omega_r t}\alpha^*)\mathbf{A}<\vec{S}>-\Gamma_{\bot}\vec{V}_z\;\;,
\label{Bloch}
\end{equation}
where $\vec{S}=\{\sigma_{-},\sigma_{+,},\sigma_{z}\}$ represents the pauli operators acting on the qubit space and where $\mathbf{A}$ and $\vec{V}$ are given by:

\begin{equation}
\mathbf{A}=\left( \begin{array}{ccc}
i\delta-\frac{1}{T_2} & 0 &i\frac{\Omega}{2} \\
0 & -i\delta-\frac{1}{T_2}& -i\frac{\Omega}{2} \\
i\Omega& -i\Omega&-\frac{1}{T_1} \end{array} \right)\;\;,
\label{A}
\end{equation}
and:
\[\vec{V}_{z}=\left(\begin{array}{c}0\\0\\1\end{array}\right).\]\;
By writing the solution in the following representation:

\begin{equation}
<\vec{S}>(t)=\sum_{n=-\infty}^{+\infty}\vec{S}_n(t)e^{-in\omega_r t}
\end{equation}
and by supposing that $\vec{S}_n(t)=\vec{S}_n$ (no time dependence), one can write the steady state bloch equation in the following way:

\begin{equation}
-\mathbf{T}\vec{S}_{n-1}+B_n\vec{S}_n-\mathbf{T}^*\vec{S}_{n+1}=\vec{V}_n\;\;,
\end{equation}

with:

\begin{equation}
\mathbf{T}=\alpha\lambda\mathbf{A}\;\;,
\end{equation}
and:
\begin{equation}
B_n=\mathbf{A}+i\omega_r n\;\;,
\end{equation}
\begin{equation}
\vec{V}_n=\delta_{n~0}\Gamma_{\bot}\vec{V}_z\;\;.
\end{equation}

To this equation introduce a parameter $m\in\mathbb{N}$ and solve iteratively in $m$ supposing, at each step, that $\vec{S}_i=0$ for $i>m$. For what follows we are interested in the solution for the terms $\vec{S}^{z}_1$ and $\vec{S}^{z}_{-1}$. They are given by the following continuous fractions:

\begin{eqnarray}\label{eq:solutions_bloch_equations}
\vec{S}_{-1}&=\frac{1}{\mathbf{K}_{-1}}\mathbf{T}^*\vec{S}_0\;\;,\\
\vec{S}_{+1}&=\frac{1}{\mathbf{K}_{+1}}\mathbf{T}\vec{S}_0\;\;,
\end{eqnarray}

with:

\begin{eqnarray}
\vec{S}_0=\frac{1}{\mathbf{R}}\vec{V}_0
\end{eqnarray}

and:

\begin{eqnarray}
\mathbf{K}_{-1}&=-\mathbf{T}~\frac{1}{\cdot}_{|_{\tiny{-2}}}~\mathbf{T}^*+B_{-1}\;\;,\\
\mathbf{K}_{+1}&=-\mathbf{T}^*~\frac{1}{\cdot^*}_{|_{\tiny{+2}}}~\mathbf{T}+B_{+1}\;\;,\\
\mathbf{R}&=\mathbf{K}_{-1}+B_0+\mathbf{K}_{+1}\;\;,
\end{eqnarray}

where $\cdot$ and $\cdot^*$ are defined recursively as:

\begin{eqnarray}
\cdot_{|_{-2}}&=-\mathbf{T}~\frac{1}{\cdot}_{|_{-3}}~\mathbf{T}^*+B_{-2}\;\;,\\
{\cdot^*}_{|_{+2}}&=-\mathbf{T}^*~\frac{1}{\cdot^*}_{|_{+3}}~\mathbf{T}+B_{+2}\;\;.
\end{eqnarray}

One can stop after a certain amount of iteration by simply substituting, instead of the previous recursive relation, the ending one:

\begin{eqnarray}
\cdot_{|_{-n}}=B_{-n}\;\;,\\
{\cdot^*}_{|_{+n}}=B_{+n}\;\;.
\end{eqnarray}

By following \cite{Rabl} the final occupation number is given by:

\begin{equation}
n_f=N_{th}\left(\frac{1}{\zeta}+\frac{1-\frac{1}{\zeta}}{1+\frac{e^{\frac{I_1\zeta}{N_{th}\eta^2}}}\zeta}\right)\;\;,
\label{Occupation}
\end{equation}

where the cooling rate $\Gamma(\alpha)$ is given by:

\begin{eqnarray}
\Gamma(\alpha)=i\lambda\left(\frac{\vec{S}^z_{1}}{\alpha}-\frac{\vec{S}^{z}_{-1}}{\alpha^*}\right)\;\;,
\end{eqnarray}

with $\vec{S}^z_1$ and $\vec{S}^z_{-1}$ are given by the solutions \ref{eq:solutions_bloch_equations} and the other terms are given by:

\begin{eqnarray}
\zeta&=\frac{\Gamma_c(0)}{\gamma}\;\;,\\
I_1&=2\int_0^\infty d\alpha~\alpha\tilde{\Gamma}_c\left(r=\frac{\alpha}{\eta}\right)\;\;,\\
\tilde{\Gamma}_c&=\frac{\Gamma_c(\alpha)}{\Gamma_c(0)}\;\;,\\
\eta&=\frac{\lambda}{\omega_r}\;\;.
\end{eqnarray}


\bibliographystyle{iopart-num}
\section*{References}
\addcontentsline{toc}{section}{References}
\bibliography{Bib}

\providecommand{\newblock}{}
\begin{thebibliography}{10}
\expandafter\ifx\csname url\endcsname\relax
  \def\url#1{{\tt #1}}\fi
\expandafter\ifx\csname urlprefix\endcsname\relax\def\urlprefix{URL }\fi
\providecommand{\eprint}[2][]{\url{#2}}

\bibitem{Bustamante}
Bustamante C, Chemla Y, Forde N and Izhaky D 2004 {\em Annu. Rev. Biochem.\/}
  {\bf 73} 705--748 ISSN {0066-4154}

\bibitem{Friedsam}
Friedsam C, Wehle A~K, Khner F and Gaub H~E 2003 {\em J. Phys.-Condens. Mat.\/}
  {\bf 15} S1709

\bibitem{Benoit}
Benoit~M G~H 2002 {\em Cells Tissues Organs\/} {\bf 172} 174

\bibitem{Rugar}
Rugar D, Budakian R, Mamin H~J and Chui B~W 2004 {\em Nature\/} {\bf 430}
  329--332

\bibitem{Ekinci}
K~L~Ekinci Y T~Yang M~L~R 2004 {\em J. Appl. Phys.\/} {\bf 95}

\bibitem{Burns1}
Burns D, Zook J, Horning R, Herb W and Guckel H 1995 {\em Sensor. actuat.
  A-Phys.\/} {\bf 48} 179 -- 186 ISSN 0924-4247

\bibitem{Schilling}
Schilling O~F 2007 {\em Braz. J. Phys.\/} {\bf 37}(2A) 425--428

\bibitem{Cleland}
Cleland A~N and Roukes M~L 1998 {\em Nature\/} {\bf 392} 160--162

\bibitem{Bocko}
Bocko M~F and Onofrio R 1996 {\em Rev. Mod. Phys.\/} {\bf 68} 755--799

\bibitem{Nakamura}
Nakamura Y, Pashkin Y~A and Tsai J~S 1999 {\em Nature\/} {\bf 398} 786--788

\bibitem{Brune}
Brune M, Hagley E, Dreyer J, Ma\^\i{}tre X, Maali A, Wunderlich C, Raimond J~M
  and Haroche S 1996 {\em Phys. Rev. Lett.\/} {\bf 77} 4887--4890

\bibitem{Clarke}
Clarke J, Cleland A~N, Devoret M~H, Esteve D and Martinis J~M 1988 {\em
  Science\/} {\bf 239} 992--997

\bibitem{Silvestrini}
Silvestrini P, Palmieri V~G, Ruggiero B and Russo M 1997 {\em Phys. Rev.
  Lett.\/} {\bf 79} 3046--3049

\bibitem{Kleckner}
Kleckner D and Bouwmeester D 2006 {\em Nature\/} {\bf 444} 75--78

\bibitem{Arcizet}
Arcizet O, Cohadon P~F, Briant T, Pinard M and Heidmann A 2006 {\em Nature\/}
  {\bf 444} 71--74

\bibitem{Gigan}
Gigan S, Bohm H~R, Paternostro M, Blaser F, Langer G, Hertzberg J~B, Schwab
  K~C, Bauerle D, Aspelmeyer M and Zeilinger A 2006 {\em Nature\/} {\bf 444}
  67--70

\bibitem{Kippenberg2}
Kippenberg T~J, Rokhsari H, Carmon T, Scherer A and Vahala K~J 2005 {\em
  Physical Review Letters\/} {\bf 95} 033901

\bibitem{Schliesser}
Schliesser A, Del'Haye P, Nooshi N, Vahala K~J and Kippenberg T~J 2006 {\em
  Physical Review Letters\/} {\bf 97} 243905

\bibitem{Schliesser2}
Thompson J~D, Zwickl B~M, Yarich A~M, Marquardt F, Girvin S~M and Harris J 2007
  {\em arXiv:0707.1724\/}

\bibitem{Corbitt}
Corbitt Tand~Chen Y~B, Innerhofer E, Muller-Ebhardt H, Ottaway D, Rehbein H,
  Sigg D, Whitcomb S, Wipf C and Mavalvala N 2007 {\em Physical Review
  Letters\/} {\bf 98} 150802

\bibitem{Chang}
Chang D~E, Regal C~A, Papp S~B, Wilson D~J, Ye J, Painter O, Kimble H~J and
  Zoller P 2010 {\em Proceedings of the National Academy of Sciences\/} {\bf
  107} 1005--1010

\bibitem{Kippenberg}
Kippenberg T~J and Vahala K~J 2007 {\em Opt. Express\/} {\bf 15} 17172--17205

\bibitem{OConnell}
O'Connell A~D, Hofheinz M, Ansmann M, Bialczak R~C, Lenander M, Lucero E,
  Neeley M, Sank D, Wang H, Weides M, Wenner J, Martinis J~M and Cleland A~N
  2010 {\em Nature\/} {\bf 464} 697--703

\bibitem{Romer}
Romer R~H 1990 {\em Eur. J. Phys.\/} {\bf 11} 103

\bibitem{Wallquist}
Jaehne K, Hammerer K and Wallquist M 2008 {\em New J. Phys.\/} {\bf 10} 095019

\bibitem{Rabl}
Rabl P 2010 {\em Phys. Rev. B\/} {\bf 82} 165320

\bibitem{Li}
Li Y, Wu L~A, Wang Y~D and Yang L~P 2011 {\em arXiv:1103.4197\/}

\bibitem{Paauw}
Paauw F~G, Fedorov A, Harmans C~J~P~M and Mooij J~E 2009 {\em Phys. Rev.
  Lett.\/} {\bf 102}(9) 090501
  \urlprefix\url{http://link.aps.org/doi/10.1103/PhysRevLett.102.090501}

\bibitem{Fedorov}
Fedorov A, Feofanov A~K, Macha P, Forn-D P, Harmans C~J~P~M and Mooij J~E 2010
  {\em Phys. Rev. Lett.\/} {\bf 105}(6) 060503
  \urlprefix\url{http://link.aps.org/doi/10.1103/PhysRevLett.105.060503}

\bibitem{McDonald}
McDonald K~T 2010 Fitzgerald's calculation of the radiation of an oscillating
  magnetic dipole
  \urlprefix\url{http://www.physics.princeton.edu/~mcdonald/examples/fitzgeral%
d.pdf}

\bibitem{Zoller}
Chang D~E, Regal C~A, Papp S~B, Wilson D~J, Ye J, Painter O, Kimble H~J and
  Zoller P 2010 {\em PNAS\/} {\bf 107} 1005--1010

\bibitem{Gustavsson}
Gustavsson S 2011 {\em arXiv:1104.5212v2\/}

\bibitem{Mamin}
Mamin H~J, Poggio M, Degen C~L and Rugar D 2007 {\em Nature\/} {\bf 2} 301--306

\end{thebibliography}

\end{document}